\documentclass[12pt,a4paper,fleqn]{article}%
\usepackage{amsmath}
\usepackage{amsfonts}

\begin{document}

\title{\textbf{Cyclic bases}\\
\textbf{of zero-curvature representations:}\\
\textbf{five illustrations to one concept}}
\author{\textsc{S.~Yu.~Sakovich}\bigskip\\
{\footnotesize Institute of Physics,
National Academy of Sciences, 220072 Minsk,
Belarus}\\{\footnotesize e-mail: saks@pisem.net}}
\date{}
\maketitle

\begin{abstract}
The paper contains five examples of using cyclic bases of zero-curvature
representations in studies of weak and strong Lax pairs, hierarchies of
evolution systems, and recursion operators.\bigskip

\end{abstract}

\section{Introduction}

Many remarkable equations of nonlinear mathematical physics can be represented
as the compatibility condition%
\begin{equation}
Z\equiv D_{t}X-D_{x}T+\left[  X,T\right]  =0 \label{zcr}%
\end{equation}
of the overdetermined linear system%
\begin{equation}
\Psi_{x}=X\Psi,\quad\Psi_{t}=T\Psi, \label{lin}%
\end{equation}
where $D_{x}$ and $D_{t}$ stand for the total derivatives; $X$ and $T$ are
$n\times n$ matrix functions of the independent variables $x$ and $t$,
dependent variables $u^{i}\left(  x,t\right)  $ and derivatives of $u^{i}$;
$\Psi\left(  x,t\right)  $ is an $n$-component column; the square brackets
denote the matrix commutator. The condition (\ref{zcr}), usually referred to
as the zero-curvature representation (ZCR), is said to represent a given
equation if any solution $u^{i}$ of the equation satisfies (\ref{zcr}). The
linear transformation%
\begin{equation}
\Psi^{\prime}\left(  x,t\right)  =G\Psi\left(  x,t\right)  \label{ltr}%
\end{equation}
of the auxiliary vector function $\Psi$ with any nondegenerate $n\times n$
matrix function $G\left(  x,t,u^{i},u_{x}^{i},\ldots\right)  $, $\det G\neq0$,
generates the gauge transformation of the matrices $X$ and $T$:%
\begin{equation}
X^{\prime}=GXG^{-1}+\left(  D_{x}G\right)  G^{-1},\quad T^{\prime}%
=GTG^{-1}+\left(  D_{t}G\right)  G^{-1}. \label{gtr}%
\end{equation}
Since the matrix $Z$ in (\ref{zcr}) is transformed under (\ref{gtr}) as%
\begin{equation}
Z^{\prime}=GZG^{-1}, \label{sim}%
\end{equation}
any two ZCRs related by a gauge transformation (\ref{gtr}) are considered as
equivalent (see, e.g., \cite{ZMNP}). The existence of such an equivalence
between ZCRs suggests that ZCRs should be studied by gauge-invariant methods.

In the present paper, we give five examples illustrating the use of the cyclic
bases of ZCRs---one of the central concepts of our work \cite{Sak1}, where a
gauge-invariant description of ZCRs of evolution equations was
proposed\footnote{A much more general (but very abstract, indeed)
gauge-invariant description, applicable to ZCRs of any non-overdetermined
systems of PDEs, was introduced by Marvan in \cite{Mar1}; later, in
\cite{Mar2}, Marvan gave explicit expressions for the objects involved.}. The
paper is organized as follows. In Section~\ref{s2}, we study a continual class
of second-order evolution equations admitting weak Lax pairs, and construct an
extension of this class to any order. In Section~\ref{s3}, we consider
removability of a parameter from a weak Lax pair of the Burgers equation, and
then, in Section~\ref{s4}, we derive the Burgers equation's recursion operator
from the cyclic basis of a strong Lax pair. In Section~\ref{s5}, we apply this
method of deriving recursion operators from ZCRs to a system of coupled KdV
equations, in order to show how the method works with systems, and then, in
Section~\ref{s6}, we use the same method again, in order to explain a strange
structure of a recursion operator of one KdV--mKdV system. Section~\ref{s7}
contains concluding remarks.

\section{Second-order evolution equations\label{s2}}

Recently, Marvan \cite{Mar3} proved that every second-order scalar evolution
equation possessing an irreducible $\mathrm{sl}_{2}$-valued ZCR can be brought
by a contact transformation into the form%
\begin{equation}
u_{t}=\beta_{x}u^{2}u_{xx}+2\beta_{xx}u^{2}u_{x}+4\beta u_{x}+\left(
\beta_{xxx}-4\beta_{x}\right)  u^{3}-4\beta_{x}u, \label{2nd}%
\end{equation}
where $\beta\left(  x,t\right)  $ is any function with $\beta_{x}\neq0$, and
then the ZCR of (\ref{2nd}) is given by (\ref{zcr}) with%
\begin{equation}
X=%
\begin{pmatrix}
\dfrac{1}{u} & 1\\
1 & -\dfrac{1}{u}%
\end{pmatrix}
, \label{mmx}%
\end{equation}%
\begin{equation}
T=%
\begin{pmatrix}
-\beta_{x}u_{x}+4\dfrac{\beta}{u}-\beta_{xx}u & 4\beta+2\beta_{x}u\\
4\beta-2\beta_{x}u & \beta_{x}u_{x}-4\dfrac{\beta}{u}+\beta_{xx}u
\end{pmatrix}
. \label{mmt}%
\end{equation}
One may wonder why the class of evolution equations (\ref{2nd}) contains the
arbitrary function $\beta\left(  x,t\right)  $. The origin of $\beta\left(
x,t\right)  $ can be revealed by the technique of cyclic bases of ZCRs
\cite{Sak1}.

Let us solve the problem of finding all the $\left(  1+1\right)  $-dimensional
scalar evolution equations%
\begin{equation}
u_{t}=f\left(  x,t,u,u_{x},\ldots,u_{x\ldots x}\right)  \label{eeq}%
\end{equation}
which admit ZCRs (\ref{zcr}) with the predetermined matrix $X$ (\ref{mmx}) and
any $2\times2$ matrices $T\left(  x,t,u,u_{x},\ldots,u_{x\ldots x}\right)  $
(traceless, without loss of generality), the highest orders of derivatives
$u_{x\ldots x}$ in $f$ and $T$ being not fixed. Since $X$ (\ref{mmx}) contains
no parameter, this problem is equivalent to finding the complete class of
evolution equations possessing weak Lax pairs with the predetermined spatial part.

First, we rewrite the ZCR of (\ref{eeq}) in its equivalent (characteristic)
form%
\begin{equation}
fC=\nabla_{x}T, \label{eqf}%
\end{equation}
where $C$ is the characteristic matrix of the ZCR (in the present case, $C$ is
simply $C=\partial X/\partial u$, because $X$ (\ref{mmx}) contains no
derivatives of $u$ \cite{Sak1,Mar2}), and the operator $\nabla_{x}$ is defined
by $\nabla_{x}Y=D_{x}Y-\left[  X,Y\right]  $ for any (here, $2\times2$) matrix
$Y$.

Second, we compute $C$, $\nabla_{x}C$, $\nabla_{x}^{2}C$ and $\nabla_{x}^{3}C$
for the matrix $X$ (\ref{mmx}), and find that the cyclic basis (i.e.~the
maximal sequence of linearly independent matrices $\nabla_{x}^{i}C$,
$i=0,1,2,\ldots$) is three-dimensional in this case, $\left\{  C,\nabla
_{x}C,\nabla_{x}^{2}C\right\}  $, and that the closure equation for the cyclic
basis is%
\begin{equation}
\nabla_{x}^{3}C=a_{1}C+a_{2}\nabla_{x}C+a_{3}\nabla_{x}^{2}C \label{clo}%
\end{equation}
with%
\begin{align}
a_{1}  &  =\frac{8u_{x}-2u_{x}^{3}}{u^{3}}-\frac{8u_{x}u_{xx}}{u^{2}}%
+\frac{12u_{x}-2u_{xxx}}{u},\nonumber\\
a_{2}  &  =\frac{4-10u_{x}^{2}}{u^{2}}-\frac{6u_{xx}}{u}+4,\quad a_{3}%
=-\frac{7u_{x}}{u}. \label{coe}%
\end{align}

Third, we decompose the unknown matrix $T$ over the cyclic basis as%
\begin{equation}
T=b_{1}C+b_{2}\nabla_{x}C+b_{3}\nabla_{x}^{2}C, \label{dec}%
\end{equation}
where $b_{1}$, $b_{2}$ and $b_{3}$ are unknown scalar functions of
$x,t,u,u_{x},\ldots,u_{x\ldots x}$. Substituting (\ref{dec}) into (\ref{eqf})
and using the closure equation (\ref{clo}) and the fact of linear independence
of $C$, $\nabla_{x}C$ and $\nabla_{x}^{2}C$, we obtain the following
relations:%
\begin{equation}
b_{2}=-D_{x}b_{3}-a_{3}b_{3},\quad b_{1}=-D_{x}b_{2}-a_{2}b_{3},\quad
f=D_{x}b_{1}+a_{1}b_{3}, \label{rel}%
\end{equation}
where the function $b_{3}$ remains arbitrary.

Fourth, we combine (\ref{coe}), (\ref{dec}) and (\ref{rel}), use the explicit
form of $C$, $\nabla_{x}C$ and $\nabla_{x}^{2}C$, and thus obtain the
following expressions for $T$ and $f$ in terms of one arbitrary function
$b_{3}=p\left(  x,t,u,u_{x},\ldots,u_{x\ldots x}\right)  $, where the order of
$u_{x\ldots x}$ is also arbitrary:%
\begin{equation}
T=%
\begin{pmatrix}
T_{11} & T_{12}\\
T_{21} & -T_{11}%
\end{pmatrix}
\label{ext}%
\end{equation}
with%
\begin{align}
T_{11}  &  =\left(  -\frac{1}{u^{2}}D_{x}^{2}+\frac{5u_{x}}{u^{3}}D_{x}%
+\frac{4-9u_{x}^{2}+3uu_{xx}}{u^{4}}\right)  p,\nonumber\\
T_{12}  &  =\left(  \frac{2}{u^{2}}D_{x}+\frac{4-6u_{x}}{u^{3}}\right)
p,\nonumber\\
T_{21}  &  =\left(  -\frac{2}{u^{2}}D_{x}+\frac{4+6u_{x}}{u^{3}}\right)  p,
\label{tij}%
\end{align}
and%
\begin{align}
f  &  =\left(  D_{x}^{3}-\frac{7u_{x}}{u}D_{x}^{2}-\frac{4+4u^{2}-24u_{x}%
^{2}+8uu_{xx}}{u^{2}}D_{x}\right. \nonumber\\
&  \left.  +\frac{16u_{x}+12u^{2}u_{x}-36u_{x}^{3}+27uu_{x}u_{xx}%
-3u^{2}u_{xxx}}{u^{3}}\right)  p. \label{exf}%
\end{align}

The problem has been solved. Every evolution equation (\ref{eeq}), which
belongs to the continual class determined by (\ref{exf}) with any function
$p(x,t,u,\allowbreak u_{x},\ldots,u_{x\ldots x})$ of any order in $u_{x\ldots
x}$, admits the ZCR (\ref{zcr}) with $X$ given by (\ref{mmx}) and $T$
determined by (\ref{ext})--(\ref{tij}). As it was shown in \cite{Sak1}, such a
continual class of evolution equations, which all possess ZCRs with a
predetermined matrix $X$, exists for every matrix $X$ containing no parameter.
Of course, the studied case of $X$ (\ref{mmx}) is not an exception.

Now we can see that the origin of the arbitrary function $\beta\left(
x,t\right)  $ in (\ref{2nd}) is the arbitrary function $p\left(
x,t,u,u_{x},\ldots,u_{x\ldots x}\right)  $ in (\ref{exf}). Indeed, if we
require that $f$ (\ref{exf}) is a function of $x,t,u,u_{x}$ and $u_{xx}$ only,
i.e.~that the order of the represented evolution equations is two, then
$p=p\left(  x,t,u\right)  $ (otherwise $f$ contains $u_{x\ldots x}$ of order
four or higher) and $\partial p/\partial u=3p/u$ (this follows from $\partial
f/\partial u_{xxx}=0$), that is%
\begin{equation}
p=\beta\left(  x,t\right)  u^{3} \label{pbu}%
\end{equation}
with any function $\beta$ ($\beta_{x}\neq0$ if we need $\partial f/\partial
u_{xx}\neq0$). Finally, the equation (\ref{pbu}) relates (\ref{exf}) with
(\ref{2nd}), as well as (\ref{ext})--(\ref{tij}) with (\ref{mmt}).

\section{Burgers equation's weak Lax pair\label{s3}}

Our next example is the ZCR (\ref{zcr}) with%
\begin{equation}
X=%
\begin{pmatrix}
\frac{1}{2}\eta & \frac{1}{4}u+\frac{1}{2}\eta\smallskip\\
\frac{1}{4}u-\frac{1}{2}\eta & -\frac{1}{2}\eta
\end{pmatrix}
, \label{wbx}%
\end{equation}%
\begin{equation}
T=%
\begin{pmatrix}
\frac{1}{4}\eta u & \frac{1}{4}u_{x}+\frac{1}{8}u^{2}+\frac{1}{4}\eta
u\smallskip\\
\frac{1}{4}u_{x}+\frac{1}{8}u^{2}-\frac{1}{4}\eta u & -\frac{1}{4}\eta u
\end{pmatrix}
, \label{wbt}%
\end{equation}
where $\eta$ is a parameter. This ZCR of the Burgers equation $u_{t}%
=u_{xx}+uu_{x}$ was introduced by Cavalcante and Tenenblat \cite{CT}. The
parameter $\eta$, however, is not an essential (`spectral') parameter: it can
be removed (`gauged out') from (\ref{wbx}) and (\ref{wbt}) by a gauge
transformation (\ref{gtr}) (therefore this ZCR is a weak Lax pair of the
Burgers equation). More precisely, it was recently shown by Marvan \cite{Mar4}
that any nonzero value of $\eta$ in these $X$ and $T$ can be changed into any
other nonzero value (e.g., into $\eta=1$) by an appropriate gauge
transformation.\footnote{The fact of removability of $\eta$ from $X$
(\ref{wbx}) was also noted in \cite{Sak1}, in slightly different notations, in
relation with ZCRs of third-order evolution equations \cite{RT}.}

Let us see how the technique of cyclic bases allows to discover, in a short
computational way, that the case of $\eta=0$ is essentially different from any
other case of nonzero $\eta$, and that the parameter $\eta$ should be expected
to be removable from this ZCR. The necessary information can be obtained from
the dimension of the cyclic basis and from the coefficients of the closure equation.

Following the scheme of Section \ref{s2}, we find the characteristic matrix
and its lower-order covariant derivatives to be%
\begin{align}
C  &  =\frac{1}{4}%
\begin{pmatrix}
0 & 1\\
1 & 0
\end{pmatrix}
,\nonumber\\
\nabla_{x}C  &  =\frac{\eta}{4}%
\begin{pmatrix}
-1 & -1\\
1 & 1
\end{pmatrix}
,\quad\nabla_{x}^{2}C=\frac{\eta u}{8}%
\begin{pmatrix}
-1 & -1\\
1 & 1
\end{pmatrix}
. \label{wbc}%
\end{align}
We see from (\ref{wbc}) that, when $\eta\neq0$, the cyclic basis is
two-dimensional, $\left\{  C,\nabla_{x}C\right\}  $, with the closure equation%
\begin{equation}
\nabla_{x}^{2}C=\frac{u}{2}\nabla_{x}C, \label{wb2}%
\end{equation}
whereas in the case of $\eta=0$ the cyclic basis is one-dimensional,
consisting of the characteristic matrix $C$ itself, with the closure equation%
\begin{equation}
\nabla_{x}C=0. \label{wb1}%
\end{equation}

Since the matrices $\nabla_{x}^{i}C$ $\left(  i=0,1,2,\ldots\right)  $ are
transformed under (\ref{gtr}) as $\nabla_{x}^{\prime i}C^{\prime}=G\left(
\nabla_{x}^{i}C\right)  G^{-1}$ \cite{Sak1,Mar2}, the dimension of the cyclic
basis and the coefficients of the closure equation are invariants with respect
to gauge transformations. Therefore no gauge transformation can change the
dimension of the cyclic basis and relate the case of $\eta\neq0$ with the case
of $\eta=0$. Moreover, the coefficients of the closure equation (\ref{wb2}) do
not contain $\eta$, and this is highly indicative of the parameter
removability\footnote{An easy way to prove this removability of $\eta$ is to
solve (\ref{gtr}) with respect to $G$ directly, taking $X^{\prime}=X|_{\eta
=1}$ and $T^{\prime}=T|_{\eta=1}$ and starting from the evident necessary
conditions $G=G\left(  x,t\right)  $ and $\left[  C,G\right]  =0$. A different
approach was used in \cite{Mar4}.}. On the contrary, if there is a parameter
in a coefficient of a closure equation of a ZCR, then the parameter is
essential, not removable from the ZCR, for the reason of the gauge invariance
of coefficients of closure equations, and Sections \ref{s4}--\ref{s6} provide
examples illustrating this phenomenon.

Using the method described in Section \ref{s2}, we can construct continual
classes of evolution equations admitting ZCRs with $X$ given by (\ref{wbx}),
separately for the two cases of the cyclic basis structure:%
\begin{equation}
u_{t}=D_{x}p\quad\mathrm{for}\quad\eta=0, \label{et0}%
\end{equation}%
\begin{equation}
u_{t}=D_{x}\left(  D_{x}+\frac{u}{2}\right)  p\quad\mathrm{for}\quad\eta\neq0,
\label{et1}%
\end{equation}
where $p$ is any function of $x,t,u,u_{x},\ldots,u_{x\ldots x}$, of any order
in $u_{x\ldots x}$. Note that the orders of the linear differential operators
in the right-hand sides of (\ref{et0}) and (\ref{et1}) are equal to the
dimensions of the corresponding cyclic bases, in accordance with the general
theorem on continual classes \cite{Sak1}.

The class (\ref{et0}) consists of evolution equations admitting the conserved
density $u$. This is a natural consequence of the cyclic basis structure in
the case of $\eta=0$. Indeed, if we consider a $\left(  1\times1\right)
$-dimensional ZCR with $X=u$ and any function $T$, then this ZCR is simply a
conservation law, $D_{t}X-D_{x}T=0$, due to $\left[  X,T\right]  =0$, and the
closure equation for $X=u$ is exactly (\ref{wb1}).

The class (\ref{et1}) is more narrow than (\ref{et0}) but much wider than
habitual discrete hierarchies of integrable equations. See \cite{Sak1} for the
results of the singularity analysis which suggest that most of equations in
such continual classes are non-integrable.

The fact that the Burgers equation $u_{t}=u_{xx}+uu_{x}$ admits a ZCR with $X$
(\ref{wbx}) is equivalent to the fact that the equation belongs to the classes
(\ref{et0}) and (\ref{et1}). It tells nothing about the Burgers equation's integrability.

\section{Burgers equation's recursion operator\label{s4}}

Now, let us study a strong Lax pair of the Burgers equation%
\begin{equation}
u_{t}=u_{xx}+2uu_{x}, \label{bur}%
\end{equation}
namely, the ZCR (\ref{zcr}) with%
\begin{equation}
X=%
\begin{pmatrix}
\frac{1}{2}\left(  u+\lambda\right)  & 0\\
1 & -\frac{1}{2}\left(  u+\lambda\right)
\end{pmatrix}
, \label{sbx}%
\end{equation}%
\begin{equation}
T=%
\begin{pmatrix}
\frac{1}{2}\left(  u_{x}+u^{2}-\lambda^{2}\right)  & 0\\
u-\lambda & -\frac{1}{2}\left(  u_{x}+u^{2}-\lambda^{2}\right)
\end{pmatrix}
, \label{sbt}%
\end{equation}
where $\lambda$ is any constant.\footnote{We are unable to indicate the first
appearance of this ZCR in the literature. In implicit form, this ZCR can be
found in \cite{AS}, among non-Abelian pseudopotentials of the Burgers
equation.} Note that these $X$ and $T$ are lower triangular matrices. However,
contrary to the approach adopted in \cite{Mar3}, we do not exclude such lower
triangular (or reducible) ZCRs from our consideration.

Computing $C$, $\nabla_{x}C$ and $\nabla_{x}^{2}C$ for the matrix $X$
(\ref{sbx}), we find that the cyclic basis is two-dimensional, $\left\{
C,\nabla_{x}C\right\}  $, with the closure equation%
\begin{equation}
\nabla_{x}^{2}C=\left(  u+\lambda\right)  \nabla_{x}C. \label{sbc}%
\end{equation}
The presence of $\lambda$ in the gauge-invariant coefficient $u+\lambda$ of
(\ref{sbc}) proves that the parameter $\lambda$ is essential, not removable by
gauge transformations. The matrix $T$ (\ref{sbt}) can be decomposed over the
cyclic basis as%
\begin{equation}
T=\left(  u_{x}+u^{2}-\lambda^{2}\right)  C+\left(  \lambda-u\right)
\nabla_{x}C. \label{edt}%
\end{equation}

Let us find all the evolution equations (\ref{eeq}) which admit ZCRs
(\ref{zcr}) with $X$ determined by (\ref{sbx}) and $T$ decomposable over the
cyclic basis as%
\begin{equation}
T=b_{1}C+b_{2}\nabla_{x}C, \label{gdt}%
\end{equation}
where $b_{1}$ and $b_{2}$ are any scalar functions of $x,t,u,u_{x}%
,\ldots,u_{x\ldots x}$ and $\lambda$. Using (\ref{eqf}), (\ref{gdt}),
(\ref{sbc}) and the fact of linear independence of $C$ and $\nabla_{x}C$, we
find that%
\begin{equation}
b_{1}=\left(  D_{x}+u+\lambda\right)  p, \label{sbb}%
\end{equation}%
\begin{equation}
f=\left(  D_{x}^{2}+\left(  u+\lambda\right)  D_{x}+u_{x}\right)  p,
\label{sbf}%
\end{equation}
where $p\left(  \lambda,x,t,u,u_{x},\ldots,u_{x\ldots x}\right)  =-b_{2}$ is
any function, of any order in $u_{x\ldots x}$. Note that the linear
differential operator in the right-hand side of (\ref{sbf}) contains the
parameter $\lambda$. If $\lambda$ is any fixed constant, then our problem is
completely solved by the expression (\ref{sbf}) for the right-hand side of
every represented equation (\ref{eeq}), and we obtain a typical continual
class. However, $\lambda$ is a free parameter, and we have to determine the
set of functions $p$ satisfying the condition $\partial f/\partial\lambda=0$
imposed on (\ref{sbf}). Here we give a detailed derivation of those admissible
$p$, in order to omit details of the same way of reasoning in Sections
\ref{s5} and \ref{s6}.

First, we introduce the operators%
\begin{equation}
M=D_{x}^{2}+uD_{x}+u_{x},\quad N=D_{x}, \label{sbo}%
\end{equation}
expand the function $p$ as%
\begin{equation}
p=p_{0}+\lambda p_{1}+\lambda^{2}p_{2}+\lambda^{3}p_{3}+\cdots, \label{exp}%
\end{equation}
where $p_{i}$ $\left(  i=0,1,2,\ldots\right)  $ are unknown functions of
$x,t,u,u_{x},\ldots,u_{x\ldots x}$, and thus rewrite (\ref{sbf}) in the
form\footnote{If a singularity is suspected in $p$ at $\lambda=0$, one may use
(\ref{mnp}) after an infinitesimal shift of $\lambda$, $\lambda\mapsto
\lambda+\epsilon,M\mapsto M-\epsilon N,N\mapsto N,\epsilon\rightarrow0$, with
no effect on the final result (\ref{bro}).}%
\begin{equation}
f=\left(  M+\lambda N\right)  \left(  p_{0}+\lambda p_{1}+\lambda^{2}%
p_{2}+\cdots\right)  . \label{mnp}%
\end{equation}

Second, we obtain from (\ref{mnp}) the expression for $f$ satisfying the
condition $\partial f/\partial\lambda=0$,%
\begin{equation}
f=Mp_{0}, \label{fmp}%
\end{equation}
as well as the recursion relation determining the functions $p_{i}$,%
\begin{equation}
Mp_{i+1}+Np_{i}=0,\quad i=0,1,2,\ldots. \label{rec}%
\end{equation}

Third, taking into account that $M$ and $N$ (\ref{sbo}) are differential
operators of orders two and one, respectively, we conclude from (\ref{rec})
that there exists a sufficiently large number $m$, such that $p_{m}$ does not
contain $u$ and derivatives of $u$. Then the recursion relation (\ref{rec})
for $i\geq m$ and the explicit expressions for $M$ and $N$ (\ref{sbo}) lead to
$p_{m}=\phi\left(  t\right)  $, with any $\phi$, and $p_{i}=0$ for $i>m$.

The problem has been solved: the right-hand side of every evolution equation
sought is determined by (\ref{fmp}), where $p_{0}$ is given by the recursion
(\ref{rec}) starting from $p_{m}=\phi\left(  t\right)  $, with any function
$\phi\left(  t\right)  $ and any integer $m\geq0$, the operators $M$ and $N$
being defined by (\ref{sbo}).

However, fourth, the result can be represented in a more compact form. Using
the operator $N^{-1}$, formally inverse\footnote{In the sense that $N^{-1}a$
denotes any function $b\left(  x,t,u,u_{x},\ldots,u_{x\ldots x}\right)  $ such
that $a=Nb$ if $b$ exists.} to $N$, we express $p_{0}$ from (\ref{rec}) as
$p_{0}=\left(  -N^{-1}M\right)  ^{m}p_{m}$, rewrite (\ref{fmp}) as $f=M\left(
-N^{-1}M\right)  ^{m}N^{-1}Np_{m}$, take into account that $Np_{m}=0$, and
thus obtain%
\begin{equation}
f=R^{m+1}0, \label{hie}%
\end{equation}
where%
\begin{equation}
R=MN^{-1}=D_{x}+u+u_{x}D_{x}^{-1}. \label{bro}%
\end{equation}

Consequently, the matrix $X$ (\ref{sbx}) is related not to a continual class
of evolution equations, as it would be if there was no essential parameter in
$X$, but to the discrete hierarchy determined by (\ref{hie}), which consists
of the evolution equations%
\begin{align}
u_{t}  &  =\phi_{1}\left(  t\right)  u_{x}+\phi_{2}\left(  t\right)  \left(
u_{xx}+2uu_{x}\right) \nonumber\\
&  +\phi_{3}\left(  t\right)  \left(  u_{xxx}+3uu_{xx}+3u_{x}^{2}+3u^{2}%
u_{x}\right)  +\cdots\label{bhi}%
\end{align}
with any functions $\phi_{1}$, $\phi_{2}$, etc. All the equations (\ref{bhi})
can be linearized by the Cole--Hopf transformation. The obtained recursion
operator $R$ (\ref{bro}), which generates the hierarchy of equations
(\ref{bhi}), is well known as a recursion operator of the Burgers equation in
the sense of higher symmetries (or generalized symmetries, or
Lie--B\"{a}cklund algebras) (see, e.g., \cite{Ibr}).

\section{Coupled KdV equations\label{s5}}

The method, used in Section \ref{s4} to derive the Burgers equation's
hierarchy and recursion operator, can be applied to systems of evolution
equations as well. Therefore our next example is%
\begin{equation}
u_{t}=u_{xxx}-12uu_{x},\quad v_{t}=4v_{xxx}-6vu_{x}-12uv_{x}. \label{gks}%
\end{equation}
These coupled KdV equations were found by G\"{u}rses and Karasu \cite{GK} as a
new system possessing a second-order recursion operator (in the sense of
higher symmetries). Later, the system (\ref{gks}) appeared in \cite{Sak2}, in
a list of coupled KdV equations that passed the Painlev\'{e} test for
integrability well, and its ZCR (\ref{zcr}) with $3\times3$ matrices $X$ and
$T$ containing an essential parameter was also found there. For what follows,
we need to know the explicit form of the matrix $X$ from \cite{Sak2},%
\begin{equation}
X=%
\begin{pmatrix}
0 & u+\sigma & 0\\
2 & 0 & 0\\
0 & v & 0
\end{pmatrix}
, \label{gkx}%
\end{equation}
where $\sigma$ is a parameter. No explicit expression for the matrix $T$ is
necessary: it is sufficient to know that $T$ can be decomposed over the cyclic basis.

In this case of a system of two equations, we have two characteristic
matrices, $C_{u}$ and $C_{v}$, which are simply $C_{u}=\partial X/\partial u$
and $C_{v}=\partial X/\partial v$ because $X$ (\ref{gkx}) contains no
derivatives of $u$ and $v$ \cite{Sak1,Mar2}. Computing $\nabla_{x}C_{u}$,
$\nabla_{x}C_{v}$, $\nabla_{x}^{2}C_{u}$, $\nabla_{x}^{2}C_{v}$ and
$\nabla_{x}^{3}C_{u}$, we find that the cyclic basis is five-dimensional,
$\left\{  C_{u},C_{v},\nabla_{x}C_{u},\nabla_{x}C_{v},\nabla_{x}^{2}%
C_{u}\right\}  $, with the two closure equations:%
\begin{align}
\nabla_{x}^{2}C_{v}  &  =2\left(  u+\sigma\right)  C_{v},\nonumber\\
\nabla_{x}^{3}C_{u}  &  =4u_{x}C_{u}+2v_{x}C_{v}+8\left(  u+\sigma\right)
\nabla_{x}C_{u}+6v\nabla_{x}C_{v}. \label{gkc}%
\end{align}
The presence of $\sigma$ in the gauge-invariant coefficients of (\ref{gkc})
proves that the parameter $\sigma$ is essential, not removable from the
studied ZCR by gauge transformations.

Next we pose the problem of finding the hierarchy of all systems%
\begin{align}
u_{t}  &  =f\left(  x,t,u,v,\ldots,u_{x\ldots x},v_{x\ldots x}\right)
,\nonumber\\
v_{t}  &  =g\left(  x,t,u,v,\ldots,u_{x\ldots x},v_{x\ldots x}\right)
\label{uvt}%
\end{align}
which admit ZCRs (\ref{zcr}) with $X$ determined by (\ref{gkx}) and $T$
decomposable over the cyclic basis as%
\begin{equation}
T=c_{1}C_{u}+c_{2}C_{v}+c_{3}\nabla_{x}C_{u}+c_{4}\nabla_{x}C_{v}+c_{5}%
\nabla_{x}^{2}C_{u}, \label{t15}%
\end{equation}
where $c_{1},\ldots,c_{5}$ are any scalar functions of $x,t,u,v,\ldots
,u_{x\ldots x},v_{x\ldots x}$ and $\sigma$. Substituting $T$ (\ref{t15}) into
the characteristic form of ZCRs%
\begin{equation}
fC_{u}+gC_{v}=\nabla_{x}T \label{cf2}%
\end{equation}
and using the closure equations (\ref{gkc}), we obtain%
\begin{align}
c_{1}  &  =-D_{x}c_{3}-8\left(  u+\sigma\right)  c_{5},\nonumber\\
c_{2}  &  =-D_{x}c_{4}-6vc_{5},\quad c_{3}=-D_{x}c_{5},\nonumber\\
f  &  =D_{x}c_{1}+4u_{x}c_{5},\quad g=D_{x}c_{2}+2\left(  u+\sigma\right)
c_{4}+2v_{x}c_{5}, \label{cfg}%
\end{align}
where the functions $c_{4}$ and $c_{5}$ remain arbitrary. The relations
(\ref{cfg}) lead to the following expression for the right-hand sides of the
represented equations (\ref{uvt}):%
\begin{equation}%
\begin{pmatrix}
f\\
g
\end{pmatrix}
=\left(  M+\lambda N\right)  p, \label{fgp}%
\end{equation}
where $\lambda=-8\sigma$, $p=\left(  c_{5},-c_{4}\right)  ^{\mathrm{T}}$
($\mathrm{T}$ denotes transposing),%
\begin{equation}
M=%
\begin{pmatrix}
D_{x}^{3}-8uD_{x}-4u_{x} & 0\\
-6vD_{x}-4v_{x} & D_{x}^{2}-2u
\end{pmatrix}
, \label{gkm}%
\end{equation}%
\begin{equation}
N=%
\begin{pmatrix}
D_{x} & 0\\
0 & \frac{1}{4}%
\end{pmatrix}
. \label{gkn}%
\end{equation}

Finally, we impose the conditions $\partial f/\partial\lambda=0$ and $\partial
g/\partial\lambda=0$ on (\ref{fgp}), in order to find the set of admissible
two-component functions $p$. Following the same way of reasoning as used in
Section \ref{s4}, we conclude that the systems (\ref{uvt}) represented by
(\ref{zcr}) with $X$ (\ref{gkx}) constitute a discrete hierarchy generated by
the recursion operator $R=MN^{-1}$ with $M$ (\ref{gkm}) and $N$ (\ref{gkn}).
Inverting the matrix differential operator $N$,%
\begin{equation}
N^{-1}=%
\begin{pmatrix}
D_{x}^{-1} & 0\\
0 & 4
\end{pmatrix}
, \label{gki}%
\end{equation}
we obtain the explicit expression for the recursion operator $R$,%
\begin{equation}
R=%
\begin{pmatrix}
D_{x}^{2}-8u-4u_{x}D_{x}^{-1} & 0\smallskip\\
-6v-4v_{x}D_{x}^{-1} & 4D_{x}^{2}-8u
\end{pmatrix}
, \label{gkr}%
\end{equation}
which coincides with the one found in \cite{GK}.

\section{Coupled KdV--mKdV equations\label{s6}}

Recently, Kersten and Krasil'shchik \cite{KK} introduced the new system of
coupled KdV--mKdV equations%
\begin{align}
u_{t}  &  =-u_{xxx}+6uu_{x}-3ww_{xxx}-3w_{x}w_{xx}+3u_{x}w^{2}+6uww_{x}%
,\nonumber\\
w_{t}  &  =-w_{xxx}+3w^{2}w_{x}+3uw_{x}+3u_{x}w \label{kks}%
\end{align}
and found its recursion operator (in the sense of higher symmetries). The
recursion operator of (\ref{kks}) has an unusual structure: its nonlocal part
consists of terms of the form $e_{1}D_{x}^{-1}$, $e_{2}D_{x}^{-1}e_{3}$ and
$e_{4}D_{x}^{-1}e_{5}D_{x}^{-1}$, and those expressions $e_{i}$ are
polynomials of $u$, $u_{x}$, $w$, $w_{x}$, $w_{xx}$, $w_{xxx}$, $y$,
$\sin\left(  2z\right)  $ and $\cos\left(  2z\right)  $, where $y$ and $z$ are
nonlocal variables,%
\begin{equation}
y:\ y_{x}=u,\quad z:\ z_{x}=w. \label{nlv}%
\end{equation}

The appearance of trigonometric functions of a nonlocal variable in the
recursion operator derived in \cite{KK} looks strange and, of course, may be
explained in various ways. One possible explanation can be obtained by the
method described in Sections \ref{s4} and \ref{s5}, which allows us to
re-derive the recursion operator of (\ref{kks}) in the quotient form
$R=MN^{-1}$, where $M$ and $N$ are $2\times2$ matrix linear differential
operators of orders four and two, respectively, whose coefficients contain
only local variables. This approach makes use of the matrix $X$ of the ZCR of
(\ref{kks}), found very recently in \cite{KSY},%
\begin{equation}
X=%
\begin{pmatrix}
\alpha & u-w^{2}+9\alpha^{2} & w\\
1 & \alpha & 0\\
0 & 6\alpha w & -2\alpha
\end{pmatrix}
, \label{kkx}%
\end{equation}
where $\alpha$ is a parameter.

Starting from the characteristic matrices $C_{u}=\partial X/\partial u$ and
$C_{v}=\partial X/\partial v$ and repeatedly applying the operator $\nabla
_{x}=D_{x}-\left[  X,\quad\right]  $, we find the cyclic basis to be
eight-dimensional,%
\begin{equation}
\left\{  C_{u},C_{v},\nabla_{x}C_{u},\nabla_{x}C_{v},\nabla_{x}^{2}%
C_{u},\nabla_{x}^{2}C_{v},\nabla_{x}^{3}C_{u},\nabla_{x}^{3}C_{v}\right\}  ,
\label{kkc}%
\end{equation}
for all nonzero values of $\alpha$. Omitting here the cumbersome closure
equations, i.e.~decompositions of $\nabla_{x}^{4}C_{u}$ and $\nabla_{x}%
^{4}C_{v}$ over (\ref{kkc}), we only note that their gauge-invariant
coefficients contain the parameter $\alpha$, which is an essential parameter
therefore. Then, the problem of finding the hierarchy of coupled evolution
equations $\left\{  u_{t}=f,\ w_{t}=h\right\}  $ represented by (\ref{zcr})
with $X$ (\ref{kkx}) leads us to%
\begin{equation}%
\begin{pmatrix}
f\\
h
\end{pmatrix}
=\left(  M+\lambda N\right)
\begin{pmatrix}
c_{7}\\
c_{8}%
\end{pmatrix}
, \label{kkh}%
\end{equation}
where $\lambda=36\alpha^{2}$; the unknown functions $c_{7}$ and $c_{8}$ of
$\lambda,x,t,u,w,\ldots,u_{x\ldots x},\allowbreak w_{x\ldots x}$ are the
coefficients at $\nabla_{x}^{3}C_{u}$ and $\nabla_{x}^{3}C_{v}$, respectively,
in the decomposition of $T$ over (\ref{kkc}); and the cumbersome differential
operators $M$ and $N$ are given below, in a simplified form. Finally, the
conditions $\partial f/\partial\lambda=\partial h/\partial\lambda=0$ lead us
to the conclusion that the represented hierarchy is generated by the recursion
operator $R=MN^{-1}$.

The second-order differential operator $N$ can be factorized as%
\begin{equation}
N=DP\ :\quad D=%
\begin{pmatrix}
D_{x} & 0\\
0 & D_{x}%
\end{pmatrix}
,\quad P=%
\begin{pmatrix}
P_{11} & P_{12}\\
P_{21} & P_{22}%
\end{pmatrix}
, \label{ndp}%
\end{equation}
where%
\begin{equation}
P_{11}=D_{x}+\frac{w_{x}}{w}+\frac{uw+2w^{3}}{w_{x}}, \label{p11}%
\end{equation}%
\begin{align}
P_{12}  &  =-2wD_{x}-12w_{x}-\frac{u_{x}}{w}\nonumber\\
&  -\frac{u^{2}+6uw^{2}+8w^{4}-2ww_{xx}}{w_{x}}+\frac{uw_{xx}}{ww_{x}},
\label{p12}%
\end{align}%
\begin{equation}
P_{21}=\frac{w^{2}}{w_{x}}, \label{p21}%
\end{equation}%
\begin{equation}
P_{22}=D_{x}-\frac{uw+4w^{3}-w_{xx}}{w_{x}}. \label{p22}%
\end{equation}

The fourth-order differential operator $M$ can be expressed as%
\begin{equation}
M=R_{\mathrm{loc}}N+Q, \label{mrn}%
\end{equation}
where $R_{\mathrm{loc}}$ turns out to be exactly the local part of the
recursion operator $R$ found in \cite{KK} (so we are on the right way),%
\begin{equation}
R_{\mathrm{loc}}=%
\begin{pmatrix}
-D_{x}^{2}+4u+w^{2} & -2wD_{x}^{2}-w_{x}D_{x}+3uw-2w_{xx}\smallskip\\
2w & -D_{x}^{2}+2u+w^{2}%
\end{pmatrix}
, \label{loc}%
\end{equation}
and the components of the first-order differential operator $Q$,%
\begin{equation}
Q=%
\begin{pmatrix}
Q_{11} & Q_{12}\\
Q_{21} & Q_{22}%
\end{pmatrix}
, \label{maq}%
\end{equation}
are determined by%
\begin{align}
Q_{11}  &  =\left(  2u_{x}+ww_{x}\right)  D_{x}+uw^{2}+w_{x}^{2}%
+ww_{xx}\nonumber\\
&  +\frac{2u_{x}w_{x}}{w}+\frac{\left(  2uw+5w^{3}\right)  u_{x}-w^{2}w_{xxx}%
}{w_{x}}, \label{q11}%
\end{align}%
\begin{align}
Q_{12}  &  =\left(  -3wu_{x}+\left(  2u-4w^{2}\right)  w_{x}-w_{xxx}\right)
D_{x}\nonumber\\
&  -u^{2}w-4uw^{3}-23u_{x}w_{x}-8ww_{x}^{2}+\left(  2u-4w^{2}\right)
w_{xx}-\frac{2u_{x}^{2}}{w}\nonumber\\
&  -\frac{\left(  2u^{2}+13uw^{2}+20w^{4}-5ww_{xx}\right)  u_{x}-\left(
uw+4w^{3}-w_{xx}\right)  w_{xxx}}{w_{x}}\nonumber\\
&  +\frac{2uu_{x}w_{xx}}{ww_{x}}, \label{q12}%
\end{align}%
\begin{equation}
Q_{21}=w_{x}D_{x}+2uw+5w^{3}+\frac{w_{x}^{2}}{w}+\frac{w^{2}u_{x}}{w_{x}},
\label{q21}%
\end{equation}%
\begin{align}
Q_{22}  &  =\left(  u_{x}+ww_{x}\right)  D_{x}-2u^{2}-13uw^{2}-20w^{4}%
-14w_{x}^{2}+5ww_{xx}\nonumber\\
&  -\frac{u_{x}w_{x}-uw_{xx}}{w}-\frac{\left(  uw+4w^{3}-w_{xx}\right)  u_{x}%
}{w_{x}}. \label{q22}%
\end{align}

Now, let us proceed to the nonlocal part $QN^{-1}$ of the recursion operator
$R=MN^{-1}$. Since $N^{-1}=P^{-1}D^{-1}$ due to (\ref{ndp}), we have to invert
the operator $P$ determined by (\ref{p11})--(\ref{p22}). Considering%
\begin{equation}
S=P^{-1}=%
\begin{pmatrix}
S_{11} & S_{12}\\
S_{21} & S_{22}%
\end{pmatrix}
\ :\quad SP=PS=%
\begin{pmatrix}
1 & 0\\
0 & 1
\end{pmatrix}
, \label{sp1}%
\end{equation}
we obtain%
\begin{align}
S_{11}  &  =-P_{21}^{-1}P_{22}S_{21},\nonumber\\
S_{12}  &  =-P_{21}^{-1}P_{22}S_{22}+P_{21}^{-1},\nonumber\\
S_{21}  &  =-\left(  P_{11}P_{21}^{-1}P_{22}-P_{12}\right)  ^{-1},\nonumber\\
S_{22}  &  =\left(  P_{11}P_{21}^{-1}P_{22}-P_{12}\right)  ^{-1}P_{11}%
P_{21}^{-1}. \label{sij}%
\end{align}
Then the relations (\ref{sij}), (\ref{p11})--(\ref{p22}) and (\ref{q11}%
)--(\ref{q22}) lead us to a cumbersome explicit expression for the nonlocal
part $QP^{-1}D^{-1}$ of the recursion operator $R$ in terms of local variables
and operators $D_{x}$, $D_{x}^{-1}$ and $K^{-1}$, where the operator $K^{-1}$
is formally inverse to the second-order differential operator%
\begin{align}
K  &  =P_{11}P_{21}^{-1}P_{22}-P_{12}\nonumber\\
&  =\frac{w_{x}}{w^{2}}D_{x}^{2}+\left(  \frac{2w_{xx}}{w^{2}}-\frac{w_{x}%
^{2}}{w^{3}}\right)  D_{x}+\left(  4w_{x}+\frac{w_{xxx}}{w^{2}}-\frac
{w_{x}w_{xx}}{w^{3}}\right)  . \label{kkk}%
\end{align}

Note, however, that no nonlocal variables have appeared as yet. The nonlocal
variable $z$ (\ref{nlv}) appears only when we make the factorization of the
operator $K$ (\ref{kkk}),%
\begin{equation}
K=\frac{\mathrm{e}^{-2\mathrm{i}z}}{w}D_{x}\frac{\mathrm{e}^{4\mathrm{i}z}}%
{w}D_{x}\mathrm{e}^{-2\mathrm{i}z}w_{x}, \label{fak}%
\end{equation}
in order to express the inverse operator $K^{-1}$ in a more conventional form,
i.e.~in terms of $D_{x}^{-1}$:%
\begin{equation}
K^{-1}=\frac{\mathrm{e}^{2\mathrm{i}z}}{w_{x}}D_{x}^{-1}\mathrm{e}%
^{-4\mathrm{i}z}wD_{x}^{-1}\mathrm{e}^{2\mathrm{i}z}w. \label{fik}%
\end{equation}
Eventually, when we rewrite the nonlocal part of $R$ in the form given in
\cite{KK}, these exponentials $\mathrm{e}^{2\mathrm{i}z}$ and $\mathrm{e}%
^{-4\mathrm{i}z}$ from $K^{-1}$ (\ref{fik}) give rise to $\sin\left(
2z\right)  $ and $\cos\left(  2z\right)  $, while the nonlocal variable $y$
(\ref{nlv}) appears simply as a result of the identity $u+yD_{x}-D_{x}y=0$ used.

\section{Conclusion\label{s7}}

Throughout the paper, we only used the cyclic bases of ZCRs as a convenient
tool, well adapted to the gauge-covariant nature of ZCRs. We believe, however,
that the cyclic bases of ZCRs can admit some interesting and useful geometric interpretation.

All the examples, studied in this paper, concerned evolution equations and
systems of evolution equations. Of course, this is not a sign of a serious
limitation for the described technique. It is well known that a very wide
class of systems of PDEs, including at least all normal systems \cite{Olv},
can be brought into the evolutionary form (see \cite{Ser} for a nice
explanation of this). There are many advantages of using the evolutionary form
of a studied system. One of them, related to ZCRs, consists in that the cyclic
basis of a ZCR is constructed of the characteristic matrices and their
successive covariant $x$-derivatives only, because the covariant
$t$-derivatives of the characteristic matrices (i.e.~$\nabla_{t}C_{u}%
=D_{t}C_{u}-\left[  T,C_{u}\right]  $, etc.) can be expressed through the
covariant $x$-derivatives in the evolutionary case \cite{Mar2}.

Further examples of using cyclic bases of ZCRs can be seen in
\cite{Sak3,KKS,Sak4}.

In \cite{Sak3}, where a ZCR with a parameter was found for the Bakirov system
that admits only one non-Lie local generalized symmetry, a gauge-invariant
coefficient of a closure equation was used as an indicator of non-removability
of the parameter, like we did it in Sections \ref{s4}--\ref{s6} of the present
paper. Note, however, that other types of gauge invariants can also be helpful
in problems of gauge non-equivalence of ZCRs and non-removability of
parameters \cite{Sak5}.

In \cite{KKS}, a recursion operator for a new integrable system of coupled KdV
equations was derived from the eight-dimensional cyclic basis of the system's
ZCR. The recursion operator, obtained in \cite{KKS}, has a strange nonlocal
part, expressed in terms of local variables and operators $D_{x}^{-1}$ and
$\left(  3D_{x}^{3}-4vD_{x}-2v_{x}\right)  ^{-1}$. Note, however, that the
nonlocal variable $w:\ v=3w_{xx}/w$ and the factorization $3D_{x}^{3}%
-4vD_{x}-2v_{x}=3w^{-2}D_{x}w^{2}D_{x}w^{2}D_{x}w^{-2}$ make it possible to
express this unusual operator $\left(  3D_{x}^{3}-4vD_{x}-2v_{x}\right)
^{-1}$ in a more conventional form, i.e.~in terms of $D_{x}^{-1}$, like we did
it by (\ref{fik}) for the Kersten--Krasil'shchik recursion operator.

In \cite{Sak4}, the technique of cyclic bases was applied to the problem of
how to distinguish the fake Lax pairs (introduced by Calogero and Nucci) from
the true ones. A clear difference was found in the structure of cyclic bases,
namely, the closure equations in the `fake' case turned out to be first-order
and typical for conservation laws, like (\ref{wb1}). Though the fake Lax pairs
are simply gauge-transformed matrix conservation laws, it still remains not so
clear what are those true Lax pairs which are generally expected to represent
only integrable (in some reasonable sense) equations. See, in this connection,
one irreducible $\mathrm{sl}_{2}$-valued ZCR with an essential parameter,
quoted in \cite{Sak1} as Example 6.


\begin{thebibliography}{99}                                                                                               %


\bibitem {ZMNP}V.E.~Zakharov, S.V.~Manakov, S.P.~Novikov and L.P.~Pitaevskii,
\textit{Theory of Solitons: the Inverse Problem Method} (Nauka, Moscow, 1980)
(in Russian).

\bibitem {Sak1}S.Yu.~Sakovich, On zero-curvature representations of evolution
equations, \textit{J.~Phys. A: Math. Gen.} \textbf{28} (1995) 2861--2869.

\bibitem {Mar1}M.~Marvan, On zero-curvature representations of partial
differential equations, \textit{Differential Geometry and Its Applications},
eds. O.~Kowalski and D.~Krupka (Silesian University, Opava, 1993) 103--122.
(ELibEMS, http://www.emis.de/proceedings/5ICDGA).

\bibitem {Mar2}M.~Marvan, A direct procedure to compute zero-curvature
representations. The case $\mathrm{sl}_{2}$, \textit{International Conference
on Secondary Calculus and Cohomological Physics}, Moscow, Russia, August
24--31, 1997, 9 pp. (ELibEMS, http://www.emis.de/proceedings/SCCP97).

\bibitem {Mar3}M.~Marvan, Scalar second-order evolution equations possessing
an irreducible $\mathrm{sl}_{2}$-valued zero-curvature representation,
\textit{J.~Phys. A: Math. Gen.} \textbf{35} (2002) 9431--9439.
(\textit{E-print} (2002) arXiv:nlin.SI/0203043).

\bibitem {CT}J.A.~Cavalcante and K.~Tenenblat, Conservation laws for nonlinear
evolution equations, \textit{J.~Math. Phys.} \textbf{29} (1988) 1044--1049.

\bibitem {Mar4}M.~Marvan, On the horizontal gauge cohomology and
non-removability of the spectral parameter, \textit{Acta Appl. Math.}
\textbf{72} (2002) 51--65. (\textit{Preprint} DIPS--2/2001, http://diffiety.ac.ru/preprint/2001/02\_01abs.htm).

\bibitem {RT}M.L.~Rabelo and K.~Tenenblat, A classification of pseudospherical
sur\-face equations of type $u_{t}=u_{xxx}+G\left(  u,u_{x},u_{xx}\right)  $,
\textit{J.~Math. Phys.} \textbf{33} (1992) 537--549.

\bibitem {AS}M.J.~Ablowitz and H.~Segur, \textit{Solitons and the Inverse
Scattering Transform} (SIAM, Philadelphia, 1981).

\bibitem {Ibr}N.H. Ibragimov, \textit{Groups of Transformations in
Mathematical Physics} (Nauka, Moscow, 1983) (in Russian).

\bibitem {GK}M.~G\"{u}rses and A.~Karasu, Integrable coupled KdV systems,
\textit{J.~Math. Phys.} \textbf{39} (1998) 2103--2111. (\textit{E-print}
(1997) arXiv:solv-int/9711015).

\bibitem {Sak2}S.Yu.~Sakovich, Coupled KdV equations of Hirota--Satsuma type,
\textit{J.~Nonlin. Math. Phys.} \textbf{6} (1999) 255--262. (\textit{E-print}
(1999) arXiv:solv-int/9901005).

\bibitem {KK}P.~Kersten and J.~Krasil'shchik, Complete integrability of the
coupled KdV--mKdV system, \textit{E-print} (2000) arXiv:nlin.SI/0010041. (To
appear in: \textit{Advanced Studies in Pure Mathematics}, Mathematical Society
of Japan).

\bibitem {KSY}A.~Karasu~(Kalkanl\i), S.Yu.~Sakovich and \'{I}.~Yurdu\c{s}en,
Integrability of Kersten--Krasil'shchik coupled KdV--mKdV equations:
singularity analysis and Lax pair, \textit{E-print} (2002) arXiv:nlin.SI/0206046.

\bibitem {Olv}P.J.~Olver, \textit{Applications of Lie Groups to Differential
Equations} (Springer, New York, 1986).

\bibitem {Ser}A.~Sergyeyev, On recursion operators and nonlocal symmetries of
evolution equations, \textit{Proc. Sem. Diff. Geom.}, ed. D.~Krupka (Silesian
University, Opava, 2000) 159--173. (\textit{E-print} (2000) arXiv:nlin.SI/0012011).

\bibitem {Sak3}S.Yu.~Sakovich, Integrability of the Bakirov system: a
zero-curvature representation, \textit{E-print} (2002) arXiv:nlin.SI/0206034.

\bibitem {KKS}A.~Karasu~(Kalkanl\i), A.~Karasu and S.Yu.~Sakovich, A strange
recursion operator for a new integrable system of coupled Korteweg--de~Vries
equations, \textit{E-print} (2002) arXiv:nlin.SI/0203036.

\bibitem {Sak4}S.Yu.~Sakovich, True and fake Lax pairs: how to distinguish
them, \textit{E-print} (2001) arXiv:nlin.SI/0112027.

\bibitem {Sak5}S.Yu.~Sakovich, On conservation laws and zero-curvature
representations of the Liouville equation, \textit{J.~Phys. A: Math. Gen.}
\textbf{27} (1994) L125--L129.
\end{thebibliography}
\end{document}